\documentclass[aps,pre,twocolumn]{revtex4-2}

\usepackage[utf8]{inputenc}
\usepackage[T1]{fontenc}
\usepackage{latexsym}
\usepackage{amsmath}
\usepackage{amsfonts}
\usepackage{amssymb}
\usepackage{amsbsy}
\usepackage{bm}
\usepackage{color}
\usepackage{xcolor}
\usepackage{graphicx}

\usepackage{float}
\usepackage{multirow}

\newcommand{\lla}{\left\langle}
\newcommand{\rra}{\right\rangle}

\newcommand{\dst}{\displaystyle}

\graphicspath{{./figures/}}

\begin{document}
\begin{titlepage}
\title {Simulating wet active polymers by multiparticle collision dynamics}
\author{Judit Clop\'es Llah\'i, Aitor Mart\'in-G\'omez, Gerhard Gompper, and Roland G. Winkler}
\affiliation{Theoretical Physics of Living Matter, Institute of Biological Information Processing and Institute for Advanced Simulation,
Forschungszentrum J\"ulich, 52425 J\"ulich, Germany }
\email{Email: g.gompper@fz-juelich.de, r.winkler@fz-juelich.de}

\begin{abstract}
The conformational and dynamical properties of active Brownian polymers embedded in a fluid depend on the nature of the driving mechanism, e.g., self-propulsion or external actuation of the monomers. Implementations of self-propelled and actuated active Brownian polymers in a multiparticle collision dynamics (MPC) fluid are presented, which capture the distinct differences between the two driving mechanisms. The active force-free nature of self-propelled monomers requires adaptations of the MPC simulation scheme, with its streaming and collision steps, where the monomer self-propulsion velocity has to be omitted in the collision step. Comparison of MPC simulation results for active  polymers in dilute solution with results of Brownian dynamics simulations accounting for hydrodynamics via the  Rotne-Prager-Yamakawa  tensor  confirm the suitability of the implementation. The polymer conformational and dynamical properties are analyzed by the static and dynamic structure factor, and the scaling behavior of the latter with respect to the wave-number  and time dependence  are discussed. The dynamic structure factor displays various activity-induced temporal regimes, depending on the considered wave number, which reflect the persistent diffusive motion of the whole polymer at small wave numbers, and the activity-enhanced internal dynamics at large wave numbers. The obtained simulation results are compared with theoretical predictions.          
                                                                                                                                        
\end{abstract}

\maketitle
\end{titlepage}

\section{Introduction}
 
A wide class of active-matter  agents exploits viscous drag with the surrounding fluid for propulsion \cite{laug:09,elge:15,shae:20}. Biological microswimmers, e.g., bacteria, algae, ciliates, and phytoplankton, are  propelled by flagella or cilia, where the frictional anisotropic of the rotating or beading flagella/cilia provides directed motion \cite{laug:09,elge:15,gomp:20}. Synthetic microswimmers  have been designed, which mimic biological microswimmer propulsion mechanisms or are powered by phoretic processes, e.g., thermophoresis or diffusiophoresis \cite{elge:15,bech:16,gomp:20,shae:20}.  Moreover, fluid-mediated interactions  determine the collective behavior  of microswimmers. An example is the motility-induced phase separation of dry active Brownian particles in two dimensions \cite{fily:12,bial:12,redn:13,marc:16.1}, which is suppressed by fluid-mediated interactions \cite{mata:14,thee:18,zant:21}.   The provided examples underline the  fundamental importance of hydrodynamic interactions by the embedding fluid for self-propelled objects and its elementary nature for locomotion, with far-reaching consequences for the structural and dynamical aspects of active matter assemblies. 

A particular kind of active matter are filaments or polymer-like structures. As is well know,  the dynamics of passive polymers in solution is determined by  fluid-mediated interactions \cite{zimm:56,doi:86,harn:96}. By now, various studies reveal the relevance of hydrodynamics \cite{bisw:17,shaf:20} and hydrodynamic interactions on the dynamics of active polymers, with a major impact even on the polymer conformations \cite{mart:19,mart:20,wink:20}.     

So far, active polymers in dilute  bulk solution have been considered mainly by applying the Rotne-Prager-Yamakawa (RPY) hydrodynamic tensor \cite{rotn:69,yama:70} to account for hydrodynamic interactions \cite{mart:19,mart:20,jian:14}. However, this approach poses severe challenges for polymers confined in complex geometries, where the hydrodynamic tensor needs to fulfill, e.g., no-slip boundary conditions on walls, and can often hardly be determined analytically. Here, other simulation approaches, which describe the fluid explicitly, are advantageous. An example is the multiparticle collision dynamics (MPC) method \cite{male:99,kapr:08,gomp:09}.    MPC  is a  coarse-grained,  particle-based mesoscale simulation approach for fluids with inherent thermal fluctuations. It has been shown that  MPC  obeys the Navier-Stokes equations with an ideal gas equation of state \cite{male:99,ihle:09}, and that it correctly captures  hydrodynamic correlations \cite{huan:12,huan:13}.  MPC has been utilized in a broad range of equilibrium, nonequilibrium, and active system simulations, in particular, applying a mechanoelastic elastic model of a microswimmer \cite{babu:12,hu:15.1,mous:20}, as well as the more generic squirmer model \cite{tao:08.1,down:09,goet:10,thee:18,zoet:18,das:19.3,qi:20,qi:20.1,clop:20}. The versatility of the MPC method facilitates a straightforward  coupling  with other simulation techniques, e.g., molecular dynamics simulations (MD) for embedded objects \cite{male:00.1,muss:05,gomp:09,zoet:18,thee:18}. Moreover, the MPC algorithm is highly parallel, and is suitable for GPU implementation with a high performance gain \cite{west:14}.

In this article, we present an implementation of an active  Brownian  polymer in MPC, where a polymer is comprised of linearly linked monomers. Two types of active polymers are considered, with self-propelled monomers (S-ABPO), and with monomers, which experience an external active force (E-ABPO). In both cases, the point-like monomers are considered as active particles, which are propelled by an active force in a direction, which changes  diffusively. The two types of monomers differ in the coupling between the active force and the fluid. A self-propelled monomer is active-force and torque free. Hence, no Stokeslet flow emerges directly by the active motion. In contrast, the external forces of E-ABPOs give rise to such a flow field.  The implementation  of E-ABPOs in MPC is straightforward and very similar to passive polymers exposed to an external  gravitational or electric field \cite{huan:10,fran:09,sing:18}. Certain aspects for E-ABPOs have already been presented before \cite{mart:20,wink:20}. However, in the case of self-propelled monomers, their equations of motion and interaction with the MPC fluid  have to be adjusted to properly account for their active-force-free character. In MPC, with its sequence of streaming and collision steps,  the active contribution to the monomer velocity has to be omitted in the collision step, and only thermal contributions have to be included to prevent the generation of a Stokeslet by active forces.

The polymer conformational and dynamical properties are analyzed in terms of the static and dynamic structure factors for the two types of polymers for various activities and persistence lengths, which illustrates the impact of activity on the polymer conformations at different intramolecular length scales. Specifically, for small wave numbers, the time dependence of the dynamic structure factor is given by the center-of-mass mean-square displacement of the active polymer. The internal polymer dynamics is visible for large wave numbers and we obtain  a stretched exponential decay of the dynamic structure factor, with an exponent close to two, the value for the active ballistic motion, as predicted by the provided analytical approximations.  

The flow field of self-propelled particles typically includes higher-order multipole contributions, e.g., force dipoles, source dipoles etc.  \cite{llop:10,yeom:14,lask:15,thee:16.1,zoet:16,wink:17,wink:18,clop:20}. In combination with polymer conformational changes, the interference of such  monomer flow fields leads to autonomous filament/polymer motion even when individual monomers are non-motile \cite{bisw:17,jaya:12,lask:15}. In the current approach, we consider point-particles and neglect the force field by active stresses, thus the monomers  corresponds to neutral squirmers \cite{clop:20}. Already the flow-fields created by intramolecular (and external) forces yield complex flow patterns --- from the level of single monomers to the full polymer ---, which  lead to particular conformational and dynamical features, such as hydrodynamically induced shrinkage of S-ABPOs \cite{mart:19}. Yet, simulations of dumbbells comprised of squirmers reveal an influence of the squirmer active stress on the dumbbell motility \cite{clop:20}. Here, further studies on polymers are desirable to resolve the influence of swimmer-specific multipoles on the polymer properties.    

The paper is organized as follows. Section \ref{sec:MPC} outlines the MPC approach. The active polymer model and its implementation in the MPC fluid are described in Sec.~\ref{sec:model_polymer_ABPO} for self-propelled as well  as externally driven   
monomers. Section~\ref{sec:conformations} presents results for the conformational properties of the polymers, and Sec.~\ref{sec:dynamics} discusses their dynamical aspects, in particular, the dynamic structure factor. Finally, Sec.~\ref{sec:summary} summarizes our findings and presents conclusions.

\section{Multiparticle Collision Dynamics Fluid} \label{sec:MPC}

The MPC fluid consists of $N$ point  particles  of mass $m$ with the positions $\bm r_i$ and  velocities $\bm v_i$ ($i=1,\dots,N$). Their dynamics proceeds in two steps --- streaming and collision. During the streaming step, particles move ballistically over a time interval $h$, which is denoted as collision time, in absence  of external forces and fields.   Hence, the positions and velocities are updated as \cite{kapr:08,gomp:09}
\begin{align}\label{eq:stream}
\bm r_i (t+h) = &  \ \bm r_i(t) + h\bm v_i (t)   ,\\
\bm v_i (t+h)= & \ \bm v_i (t) .
\end{align}
The presence of external fields modifies the dynamics and the equations of motion may have to be solved by, e.g., the velocity Verlet algorithm \cite{fran:09,sing:18,mart:20}.
In the collision step, the system is partitioned into a lattice of cubic cells of length $a$, which define the local interaction environment. Coupling and linear momentum exchange between the $N_c$ particles of a collision cell  is achieved  by a rotation of the relative velocities
$\Delta \bm v_i(t) = \bm v_i (t)   - \bm v_{cm}(t)$,  with respect to the center-of-mass velocity $\bm v_{cm} (t) = \sum_i \bm v_i (t)/N_c$ of the cell, around a randomly oriented axis by an angle $\alpha$ (MPC-SRD) \cite{gomp:09}. The orientation of the rotation axis is chosen randomly and independently for every cell and collision step.
This yields the final velocities 
\begin{equation}
\bm v_i (t+h)= \bm v_{cm}(t+h) + \mathrm{\bf R}(\alpha) \Delta  \bm v_i (t+h)  ,  
\end{equation}
with   $\mathbf{\bf R}(\alpha)$ the rotation matrix \cite{huan:10.1}. This scheme conserves momentum on the collision cell level. 
However,  angular momentum is not conserved, which  does not affect the polymer dynamics, because the monomers are treated as point particles and, hence, possess no rotational degrees of freedom. Angular momentum conserving algorithms are provided in Refs.~\cite{nogu:07,thee:18,yang:15}.  
A constant local temperature is maintained by a collision cell-based, Maxwellian thermostat, where the relative velocities of the particles in a collision cell are scaled according to the Maxwell–Boltzmann scaling (MBS) method \cite{huan:10.1}. By  the construction of the algorithm for S-ABPOs, the thermostat only affects the thermal velocities of the monomers and the MPC particles, and not the active velocity. In case of E-ABPOs the relative velocities, including contributions by activity velocities, are scaled, as for any other external force.     
Discretization in collision cells implies violation of Galilean invariance, which is reestablished by a random shift of the collision-cell lattice after every streaming step \cite{ihle:03,gomp:09}.

\section{Active Brownian Polymers} \label{sec:model_polymer_ABPO}

\subsection{Polymer model}

We consider linear semiflexible polymers composed  of $N_m$ active point-like monomers of mass $M$, positions $\bm r_k$, and velocities $\bm v_k$ ($k=1,\ldots,N_m$), which are connected by harmonic bonds with the potential $U_l$. Bending stiffness is taken into account by restrictions of bond orientations via the bending potential $U_b$, and excluded-volume interactions by a truncated purely repulsive Lennard-Jones potential, where \cite{mart:20} 
\begin{align} \label{eq:pot_bond}
U_{l} = & \ \frac{\kappa_{l}}{2} \sum^{N_m}_{k=2} \left(|\bm{R}_k|  - \ l \right)^2 , \\ \label{eq:pot_bend}
U_{b} = & \ \frac{{\tilde \kappa}_{b}}{2} \sum^{N_m -1}_{k=2} \left(   \bm{R}_{k+1} -  \bm{R}_{k}  \right)^2 , \\ \label{eq:ev}
U_{LJ} = & \  4 \epsilon \dst \sum_{k<k'} \left[
\left( \dst \frac{\dst \sigma}{\dst  r_{kk'}} \right)^{12} - \left( \dst \frac{\dst \sigma}{\dst r_{kk'}} \right)^6 + \dst \frac{1}{4}\right]
\Theta(\sqrt[6]{2} \sigma -r_{kk'})  , 
\end{align}
with the  bond vector $\bm R_{k+1} = \bm r_{k+1} - \bm r_k$  of equilibrium length $l$. $\kappa_l$ and $\tilde \kappa_b$ characterize the strength of the bond and bending potential.  The  vector $\bm{r}_{kk'}=\bm{r}_{k}-\bm{r}_{k'}$ is the vector between monomers $k$ and $k'$, $r_{kk'} = |\bm r_{kk'}|$, and $\Theta(x)$ is the Heaviside function. The energy $\epsilon$ measures the strength of the repulsive potential and $\sigma$ defines  the particle diameter.  

The bending rigidity, $\tilde \kappa_b$, is related to the  polymer persistence length $l_p = 1/(2p)$ by 
\begin{equation}\label{eq:pL}
pL = N_m \frac{{\kappa}_{b} \left( 1 - \coth \left({\kappa}_{b} \right) \right) +1}{
	{\kappa}_{b} \left( 1 + \coth \left( {\kappa}_{b} \right) \right) -1} \ ,
\end{equation}
where $\kappa_b =\tilde \kappa_{b} l^2 /k_BT$ is the scaled bending rigidity, $k_BT$ the thermal energy with $k_B$ the Boltzmann constant and $T$ the temperature,  and $L = (N_m-1)l$ the polymer length. We will use $\kappa_b$ to characterize the polymer stiffness in the following.

\subsection{Active Brownian particle monomers} 

The activity of our monomers is captured in an active Brownian particle-type (ABP) manner \cite{bech:16,elge:15}. This implies an active translational motion of the point particles with a velocity $\bm v_k^a = v_0 \bm e_k$ of magnitude $v_0$ and direction $\bm e_k$, and a change of $\bm e_k$ in a diffusive manner according to the equations of motion
\begin{align} \label{eq:orient} 
\frac{d}{dt} \bm e_k(t) = \bm \Xi_k (t) \times \bm e _k(t) ,
\end{align}
where $\bm \Xi_k (t)$ is a Gaussian and Markovian stochastic process with zero mean and the second moments  
\begin{align}
\lla \Xi_{\alpha k} (t) \Xi_{\beta k'} (t') \rra = 2 D_R \delta_{\alpha \beta} \delta_{k k'} \delta(t-t') .
\end{align}
Here, $D_R$ is the rotational diffusion coefficient of a spherical colloid and $\alpha, \ \beta \in \{x,y,z\}$ refer to the axis of the Cartesian reference frame. Equation~\eqref{eq:orient} yields the correlation function 
\begin{align} \label{eq:prop_corr}
\lla \bm e_k (t) \cdot \bm e_k(0) \rra = e^{-2 D_R t}
\end{align}
in three dimensions. 
We like to emphasize that the active velocity is an independent quantity, neither affected by intrapolymer  forces nor hydrodynamic interactions. 

The activity is characterized by the P\'eclet number 
\begin{align} \label{eq:peclet}
Pe = \frac{v_0}{l D_R} .
\end{align}
As for hard spheres in a fluid, we fix the ratio between the translational, $D_T= k_BT/\gamma_T$, and rotational diffusion coefficient 
\begin{align}
\Delta = \frac{D_T}{(2R_h)^2 D_R} = \frac{1}{3} , 
\end{align}
where $R_h$ is the hydrodynamic radius of a monomer, which defines the translational friction coefficient $\gamma_T$ for a given $D_R$.

\subsection{Dynamics of polymers with self-propelled ABP monomers in MPC fluid (S-ABPO)}

An essential aspect of a self-propelled monomer embedded in a fluid is that active forces do not give rise to a Stokeslet, i.e., a monomer is active-force free. However, monomers experience forces due to intramolecular interactions, e.g., the forces of Eqs.~\eqref{eq:pot_bond} - \eqref{eq:ev}. Within a coarse-grained description, we are free to model the active process such that these aspects are correctly captured. To account for  non-active forces, equations of motion for  auxiliary particle  positions $\tilde {\bm r}_k$ are introduced as 
\begin{align} \label{eq:eom_self}
M \frac{d^2}{dt^2} \tilde{\bm r}_k(t) = \bm F_k(t)  ,
\end{align}  
with forces, $\bm F_k(t)$,  from the potentials \eqref{eq:pot_bond} - \eqref{eq:ev}.
The solution of these equations yields the velocities $\tilde{\bm v}_k= d \tilde{\bm r}_k/dt$. Taking the active process into account, the actual particle velocities are 
\begin{align}
\frac{d}{dt} \bm r_k \equiv \bm v_k (t) = \tilde{\bm v}_k (t) + \bm v_k^a(t) , 
\end{align}
and the final positions $\bm r_k$ follow by integration.  Using a velocity  Verlet-type approach, we obtain the following scheme for the integration of the equations of motion: 
\begin{align}
\tilde {\bm r}_k(t+\Delta t) = & \ \bm r_k (t) + \Delta t  \tilde{\bm v}_k(t)   + \frac{\Delta t^2}{2M} \bm F_k(t)\ , \\
\bm r_k(t+\Delta t) = & \ \tilde {\bm r}_k(t+\Delta t)  +  \Delta t  \bm v_k^a (t) \ , \\
\tilde {\bm v}_k(t+\Delta t)= & \  \tilde{\bm v}_k(t) + \frac{\Delta t}{2M} \left[ \bm F_k(t) + \bm F_k (t+ \Delta t) \right] \ , \\ \label{eq:velo_tilde} 
 \bm v_k(t+\Delta t)= & \ \tilde {\bm v}_k(t+\Delta t) + \bm v_k^a(t+\Delta t) \ ,
\end{align}
with the integration time step $\Delta t$ and $\bm F_k(t) = \bm F_k({\bm r(t)})$. The scheme for the integration of Eq.~\eqref{eq:orient} is described in Ref.~\cite{wink:15}. 

Coupling of the polymer with the MPC fluid is established by incorporation  of the monomers in the collision step. Here, the monomers are sorted into collision cells according to their positions $\bm r_k(t+h)$, but only the velocities $\tilde{\bm v}_k$ are taken into account in the collision to ensure active force-free motion. Hence, the monomer velocities after collision are
\begin{align} \label{eq:coll_mon}
\tilde{\bm v}_k (t+h) =  \tilde{\bm v}_{cm}(t+h) + \mathrm{\bf R}(\alpha) \left[ \tilde {\bm v}_k (t+h)   - \tilde{\bm v}_{cm}(t+h) \right] \ ,
\end{align} 
with 
\begin{align} \label{eq:com_mon}
\tilde{\bm v}_{cm} = \frac{\sum_{i=1}^{N_c} m \bm v_i + \sum_{k=1}^{N_c^m} M \tilde{\bm v}_k}{mN_c + M N_c^m} , 
\end{align}
where $N_c^m$ is the number of monomers in the collision cell of particle $k$.

\subsection{Dynamics of polymers with externally driven monomers in MPC fluid (E-ABPO)}

In case of an externally driven  monomer,   the active force is $\bm F_k^a = \gamma_T \bm v_k^a = \gamma_T v_0 \bm e_k$, where $\gamma_T$ is the translational friction coefficient, and its  equation of motion becomes
\begin{align}
M \frac{d^2}{dt^2} \bm r_k  = \bm F_k(t) + \bm F_k^a (t), 
\end{align}
 with the force $\bm F_k$ of Eq.~\eqref{eq:eom_self}.  The external forces induce an overall fluid flow, because their sum, 
\begin{align}
\bm F^a(t) =  \sum_{k=1}^{N_m} (\bm F_k(t) + \bm F_k^a(t))  =  \sum_{k=1}^{N_m} \gamma_T v_0 \bm e_k(t) ,
\end{align}
is non-zero, which leads to  fluid backflow in confined systems. In a system with periodic boundary conditions, this is achieved by a backflow force applied on the MPC fluid and a force applied on the monomers, such that the total momentum of the system is zero \cite{mart:20}.  
The  backflow force, $\bm F_k^b(t)$,  on a monomer is
\begin{align}
\bm F_k^b = - \frac{M}{mN + M N_m} \bm F^a .
\end{align}
The monomer  dynamics  is then described by the equations of motion
\begin{align} \label{eq:eom_mon_external}
M \frac{d^2}{dt^2} \bm r_k = \bm F_k + \bm F_k^a + \bm F_k^b ,
\end{align}
which is solved by applying the velocity-Verlet algorithm \cite{alle:87,mart:20}.

Under the assumption of a very slow change of the $\bm e_i$ for a small diffusion coefficient $D_R$ during a collision-time interval $h$, the MPC fluid particle velocities and positions after streaming are 
\begin{align}
\bm v_i(t+h) = & \ \bm v_i(t) - \frac{h}{mN + M N_m} \bm F^a(t) , \\ 
\bm r_i(t+h) = & \ \bm r_i(t) + h \bm v_i(t) - \frac{h^2}{2(mN + M N_m)} \bm F^a(t)  
\end{align} 
in presence of the backflow force $m \bm F^b/M$.

\begin{figure}[t]
	\begin{center}
		\includegraphics[width=1.0\columnwidth]{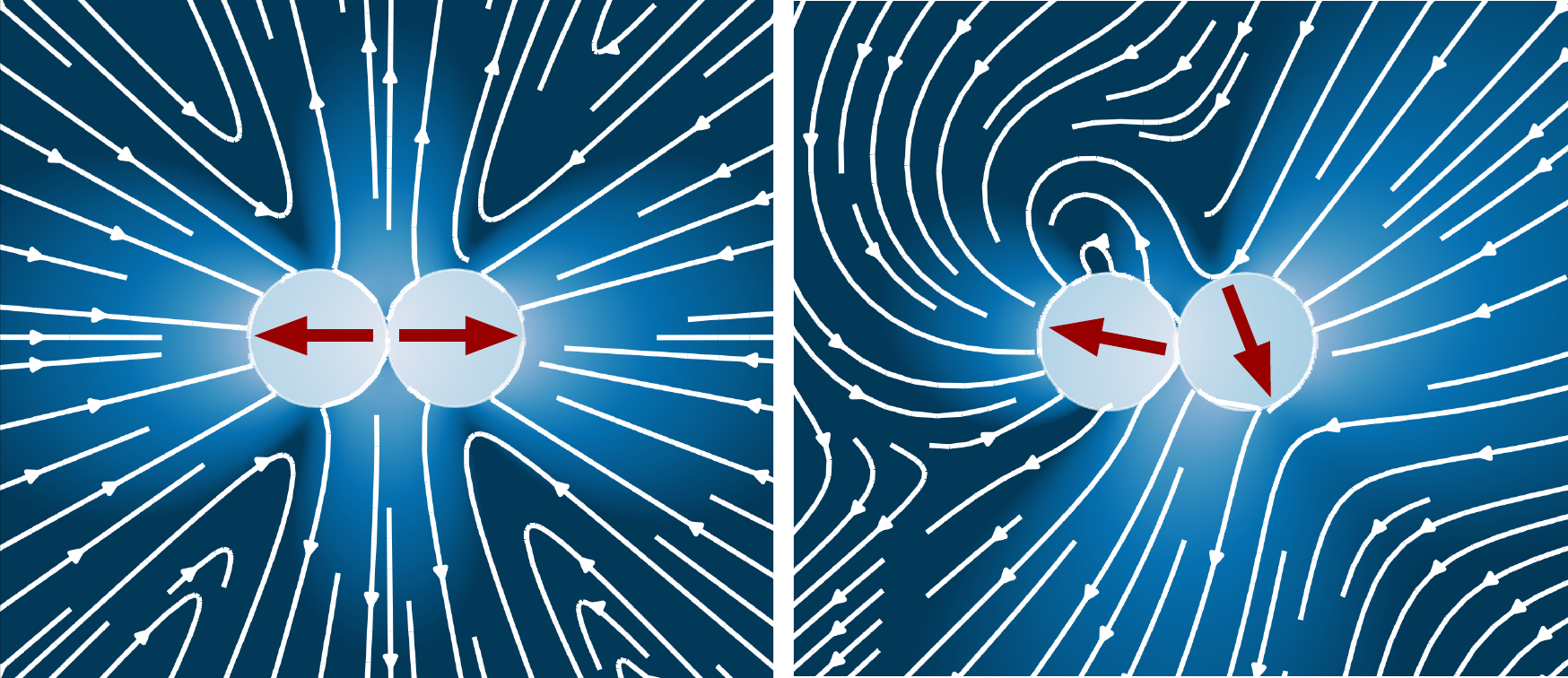}
	\end{center}
	\caption{Sketch illustrating the differences in the flow fields of self-propelled  (left) (S-ABPOs) and externally driven dumbbells (right) (E-ABPOs)  highlighting the distinct features of the propulsion mechanisms (arbitrary units). For S-ABPOs, the self-propulsion velocity does not contribute to the flow field and it is determined by bond forces, indicated by the arrows, only, i.e., a force-dipole flow field appears by the Stokeslets  of the two monomers. The flow field of an E-ABPO  comprises the flow field of the S-ABPO  as well as contributions  from the Stokeslets flows of the active monomers with their directions indicated by the arrows.  }   \label{fig:sketch}
\end{figure}

MPC collisions involving monomers implies the rotation of their velocities according to Eq.~\eqref{eq:coll_mon}, however, with the velocities  following as solution of Eq.~\eqref{eq:eom_mon_external}, i.e., the actual velocities, $\bm v_k(t+k)$, after  streaming  are used rather than $\tilde{\bm v}_k (t) $. Thus, a  Stokeslet flow field by the active forces appears  for every monomer, in addition to the flow field by passive forces. Figure~\ref{fig:sketch} illustrates the differences in the emerging flow fields for S-ABPOs and E-ABPOs.

\subsection{Parameters}

We measure length in units of the bond length $l$,  energy in units of the thermal energy $k_BT$, and time in units  of $\tau=\sqrt{ml^2/k_BT}$. The MPC particle mass is set to  $m=1$, the collision cell size to $a=l$, the rotation angle to $\alpha=130^{\circ}$, and the collision time step to $h=0.01 \tau$. MPC is an ideal gas with the isothermal velocity of sound $c_T = \sqrt{k_BT/m}$, which is unity in  units of the simulation. 

In case of the E-ABPO, we choose the average MPC particle number per collision cell $\langle N_c \rangle =10$, which yields the viscosity $\eta = 82.1 \sqrt{m k_BT/a^4}$ according to the theoretical formula \cite{huan:15}. The rotational diffusion coefficient is set to $D_R=10^{-2}/\tau$ and the length of the cubic simulation box to $L_b= 60 a$. 

For the S-ABPO, the  average MPC particle number per collision cell is $\langle N_c \rangle =50$, which yields the viscosity $\eta = 447.4 \sqrt{m k_BT/a^4}$ \cite{huan:15}. The rotational diffusion coefficient is set to $D_R=3 \times 10^{-3}/ \tau$ and the length of the qubic simulation box to $L_b= 50 a$. 

We consider a polymer with $N_m=50$ monomers, i.e., its length is $L=(N_m-1) l$.
The monomer mass is set to $M=m \langle N_c \rangle$ to achieve a suitable hydrodynamic coupling with the MPC fluid \cite{ripo:05}. The small collision time step implies the monomer hydrodynamic radius $R_h=0.3a$ \cite{kowa:13}.
The time step for solving the monomer equations of motion is $\Delta t =  10^{-3} \tau = h/10$. The coefficient  $\kappa_{l}$  for the bond strength is adjusted according to the applied active force strength, in order to avoid bond stretching with increasing activity. By choosing $\kappa_{l} l^2 /k_BT = (1  +  2 Pe ) \times 10^3$,  bond-length variations are smaller than $3 \%$ of the equilibrium value $l$. For semiflexible polymers, we consider the bending rigidity $\kappa_b=10$ and $\kappa_b = 10^3$, which yields via Eq.~\eqref{eq:pL}  $pL = 2.6 $ and $pL = 2.5 \times 10^{-2}$ or the persistence lengths $l_p = 9.4l$ and $l_p = 980 l$ ($l_p=1/(2p)$), respectively.

\section{Conformational properties} \label{sec:conformations}

\subsection{Mean-square end-to-end distance}

\begin{figure}[t]
	\begin{center}
		\includegraphics[width=1.0\columnwidth]{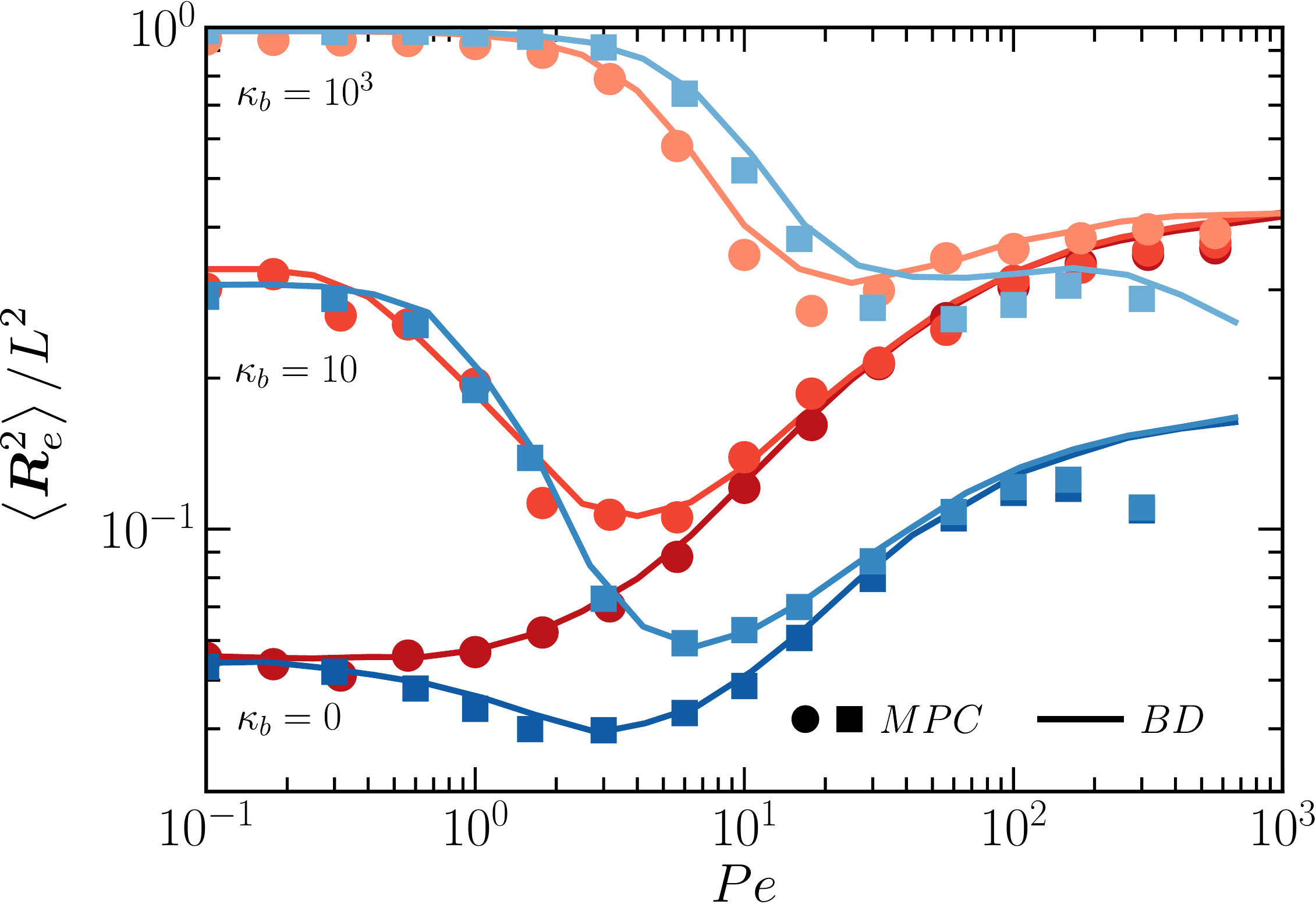}
	\end{center}
	\caption{Polymer mean-square end-to-end distance as a function of the P\'eclet number of semiflexible polymers with $N_m=50$ ($L=49l$) monomers for  $\kappa_b=0,$  $10,$ and $10^3$  (dark to bright, bottom to top). Results of  MPC  simulations (squares, bullets) and  BD simulations with the RPY hydrodynamic tensor (solid lines) are shown for  S-ABPOs (blue, squares) and   E-ABPOs (red, bullets). } \label{fig:end_SP_ED}
\end{figure}

We characterize the polymer conformational properties by the mean-square end-to-end distance $\langle \bm R_e^2 \rangle = \langle (\bm r_N - \bm r_1)^2 \rangle$. Results for E-ABPOs and S-ABPOs are presented in Fig.~\ref{fig:end_SP_ED} for  simulations with the MPC implementation (bullets, squares) as well as for Brownian dynamics (BD) simulations, where fluid interactions are taken into account by the Rotne-Prager-Yamakawa  (RPY) hydrodynamic  tensor (solid lines) \cite{mart:19,mart:20}.  As already shown previously, semiflexible polymers shrink for moderate P\'eclet numbers and swell for high $Pe$ independent of the nature of the propulsion mechanism  \cite{eise:16,mart:19,mart:20,wink:20}. However, in presence of hydrodynamic interactions even flexible polymers with self-propelled monomers shrink for $0.1 < Pe \lesssim 10$, in contrast to externally driven ones, which monotonically swell with increasing P\'eclet number \cite{mart:19,mart:20}.  Moreover, the asymptotic value of $\langle \bm R_e^2 \rangle$ of  S-ABPOs is smaller than that of E-ABPOs \cite{mart:19,wink:20}, reflecting the differences in the hydrodynamic coupling.  The hydrodynamic contribution to $\langle \bm R_e^2 \rangle$ for E-ABPOs decreases with increasing swelling, and the asymptotic value in the limit $Pe \to \infty$ agrees with that of a dry active Brownian polymer (D-ABPO), i.e., a polymer without hydrodynamic interactions \cite{mart:20,wink:20}.

The mean-square end-to-end distances obtained by the two approaches (MPC, BD) agree very well with each other. However, the results of the  MPC approach deviate  from the RPY tensor simulations for $Pe \gtrsim 10^2$, especially for the S-ABPO implementation.  This is attributed to the compressibility of the MPC fluid. The diffusive transport of vorticity (shear waves) has to be faster than (active) diffusion of the polymer to establish proper hydrodynamic interactions \cite{huan:12,huan:13}, which is achieved by avoiding large velocities $v_0$. However, to reach  large $Pe$ at the same time, a smaller $D_R$ has to be chosen. Moreover, the oscillatory Reynolds number $Re_T=\tau_{\nu}/\tilde \tau_1 $ has to obey $Re_T \ll 1$ \cite{thee:14,laug:11}, where $\tau_{\nu} $ is the viscous time scale and  $\tilde \tau_1$ the longest polymer relaxation time in presence of hydrodynamic interactions \cite{huan:12,huan:13,mart:19,mart:20,wink:20}.   
Since $\tilde \tau_1 \sim \eta$, but decreases with in increasing P\'eclet number, a large viscosity is needed to  ensure low-Reynolds number hydrodynamics, especially for S-ABPOs.  Our simulations of E-ABPOs suggest that here the requirements are  less stringent  and smaller viscosities, hence shorter relaxation times, already yield accurate results. This is related to the different coupling of the active monomer and fluid motion by inclusion of the active velocity in the collision step. Furthermore, E-ABPOs behave as free-draining polymers for $Pe \gg 1$ and asymptotically approach the mean-square end-to-end curves of D-ABPOs \cite{wink:20}.

 \begin{figure}[t]
 	\begin{center}
 		\includegraphics[width=1.0\columnwidth]{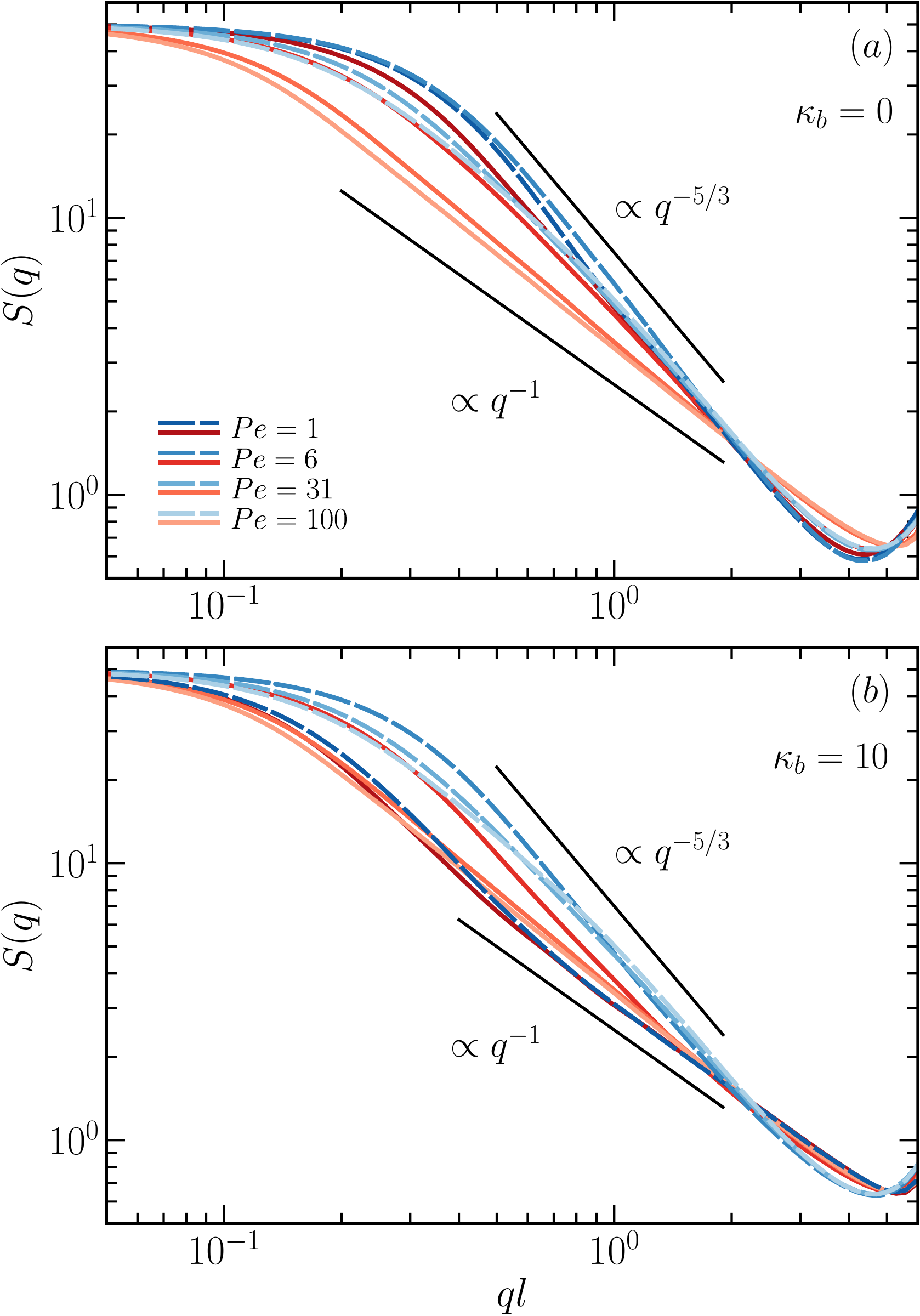}
 	\end{center}
 	\caption{Static structure factors of S-ABPOs (dashed blue) and E-ABPOs (solid red)  as a function of the wave vector $q$ for the polymer stiffness (a) $\kappa_b =0$ and (b) $\kappa_b = 10$, and the indicated P\'eclet numbers $Pe=1, 5, 31$ and $100$ (dark to bright). The short black lines indicate power-law regimes for self-avoiding, $q^{-5/3}$,  and rodlike polymers, $q^{-1}$.} \label{fig:struct_kb}
 \end{figure}

\subsection{Static structure factor}

In order to provide a more detailed insight into the active polymer conformational properties on the various length scales, we determine the static structure factor
\begin{align} \label{eq:stat_struct_fact}
S(\bm q)  = \frac{1}{N_m} \sum_{k=1}^{N_m} \sum_{n=1}^{N_m} \lla e^{-i\bm q \cdot \bm (\bm r_k - \bm r_{n})} \rra ,
\end{align} 
where $\bm q$ is the wave vector. Figure~\ref{fig:struct_kb} illustrates the strong dependence of $S(\bm q)$ on activity and the driving mechanism. Sufficiently long flexible self-avoiding passive polymers exhibit the power-law relation $S(\bm q) \sim q^{-5/3}$ ($q= |\bm q|$) over a polymer-length dependent $q$ range \cite{doi:86,dege:71}. Analogously, rodlike polymers show the power-law relation $S(\bm q) \sim q^{-1}$ \cite{doi:86}. Dependent on the P\'eclet number, both regimes are present in Fig.~\ref{fig:struct_kb}. 

The structure-factor curves of flexible E-ABPOs show a continuous change of $S(\bm q)$ from the (approximate) $q^{-5/3}$ power-law decay  at $Pe=0$ to  $q^{-1}$ for $Pe =100$, as a consequence of the swelling of the polymer, as displayed in Fig.~\ref{fig:end_SP_ED}(a), and reflected by the shift of the crossover from the small $q$-plateau  to the power-law regime,  and an associated stiffening. An explanation of  activity  on $S(\bm q)$ is provided in Sec.~\ref{sec:struct_anal} below. The structure factors of  S-ABPOs exhibit a similar behavior, except for $Pe=5$, corresponding approximately to the minimum of $\langle \bm R_e^2 \rangle$ in Fig.~\ref{fig:end_SP_ED}. The latter $S(\bm q)$ curve is shifted to larger $q$ for $ql \lesssim 2$ compared to that for $Pe=1$, but drops faster for $ql \gtrsim 1$ reflecting the polymer shrinkage  and a smaller radius of gyration  (Fig.~\ref{fig:end_SP_ED}). The polymer swelling with increasing $Pe$ increases the slope toward $-1$. Noteworthy, for $Pe=100$, the structure factor curve is flatter in the vicinity of $ql = 0.5$ and becomes steeper for larger $q$ values. Hence, the swollen polymer is stiffer on larger scales, but nearly as flexible as a passive polymer on more local scales. 

The variation in the slope as function of $Pe$ is more pronounced for stiffer polymers.  The structure factors for $\kappa =10$ and $Pe =1$ of E-ABPOs and S-ABPOs  are very similar, specifically, they reflect the intrinsic stiffness at scales  $ql >0.5$ with the dependence $S(\bm q) \sim q^{-1}$   (Fig.~\ref{fig:end_SP_ED}(b)). The activity-induced shrinkage, with a minimum at $Pe \approx 6$,  and correspondingly smaller radius of gyration, implies a steeper slope with a power-law $S(\bm q) \sim q^{-5/3}$. The swelling of the polymers for large P\'eclet numbers  causes also a stiffening  of the semiflexible polymer.  As  shown in Fig.~\ref{fig:end_SP_ED}, the swelling for $Pe \gtrsim 30$ is independent of stiffness, hence, the respective curves ($Pe=31, 100$) in Figs.~\ref{fig:struct_kb}(a), (b) are  (nearly) identical. For even stiffer polymers, $S(q)$ displays the  power law $q^{-1}$ as for a rodlike object, but for rather different physical mechanism --- at $Pe = 1$ (and similarly for $Pe <1$) due to intrinsic stiffness and  at $Pe \gtrsim 30$ due to activity.

\subsection{Analytical considerations} \label{sec:struct_anal}

Qualitatively, the activity-induced change in the structure factor can be explained by the dry  active Brownian  polymer model (D-ABPO) \cite{eise:16,wink:20}.  Adopting the Gaussian polymer model in the continuum limit \cite{eise:16,mart:19,mart:20,wink:20}, the mean-square distance $\Delta \bm r^2(s,s') = \langle (\bm r(s) - \bm r(s'))^2 \rangle$ between two points at $\bm r(s)$ and $\bm r(s')$ on the polymer, $0\leq s \leq L$ is the contour coordinate of the polymer of length $L$, of a flexible polymer ($pL \gg 1$) is given by 
\begin{widetext}
\begin{align}
\lla \Delta \bm r^2(s,s') \rra =  \left( 1 + \frac{Pe^2}{6 \Delta}\right) \frac{|s-s'|}{p \mu} + \frac{Pe^2L^2}{2} \sqrt{\frac{1}{6 pL \mu N^3 \Delta}} \bigg(  & \frac{\cosh(\beta (s+s')/L) ( 1- \cosh(\beta |s-s'|/L))}{\sinh(\beta)}  \\ \nonumber 
& \left. + \frac{\cosh(\beta) (\cosh(\beta |s-s'|/L) -1)}{\sinh(\beta)} - \sinh(\beta |s-s'|/L) .
\right) , 
\end{align}
\end{widetext} 
where $\beta = \sqrt{2 N^3/(3 pL \mu \Delta)}$ and $N$ the number of active sites.  The stretching coefficient $\mu= \mu(pL,Pe)$ accounts for the inextensibility of a polymer and depends nonlinearly  on the activity  \cite{eise:16,eise:18.1}.
In contrast to a passive polymer, $\Delta \bm r^2(s,s') $ is not only a function of $s-s'$, but depends also on $s+s'$.
Taylor expansion for $(s-s')/L \ll 1$ yields
\begin{align} \label{eq:dist_approx}
\lla \Delta \bm r^2(s,s') \rra  \approx   \frac{Pe^2 \beta^2}{4} \sqrt{\frac{1}{6 p L \mu N^3 \Delta}}  (s-s')^2 
\end{align}
for $Pe \gg 1$ and $(s+s')/L <1$. Hence, $\Delta \bm r^2(s,s') $ depends quadratically on the distance $s-s'$ due to the active forces.   

The Gaussian nature of the polymer implies that the structure factor is given by 
\begin{align}
S(\bm q) =  \frac{1}{L^2} \int_0^L \int_0^L  \exp \left(- \frac{\bm q^2}{6} \langle  \Delta \bm r^2(s,s') \rangle \right)  ds ds'  .
\end{align}
For $qL \gg1$, insertion of Eq.~\eqref{eq:dist_approx} yields  $S(\bm q) \sim \mu^{4/3}/(q Pe)$.  Hence, we obtain the dependence $S(\bm q) \sim 1/q$ for $Pe \gg 1$ as for rodlike polymers. Moreover, with $\mu \sim Pe^2 N/(pL)$ \cite{eise:16,wink:20}, $S(\bm q)$ becomes independent of $Pe$, as is reflected in Fig.~\ref{fig:struct_kb} for S-ABPOs and E-ABPOs.

\section{Dynamical properties} \label{sec:dynamics}

\subsection{Center-of-mass mean-square displacement} 

The center-of-mass mean-square displacement (CM-MSD), $\langle\Delta \bm r_{cm}^2 (t) \rangle  = \langle (\bm r_{cm} (t) - \bm r_{cm}(0))^2 \rangle$, of D-ABPOs can easily be calculated, since $\langle\Delta \bm r_{cm}^2 (t) \rangle$ is independent of intramolecular forces, and only external noise and active forces contribute \cite{eise:16,wink:20}.  The presence of hydrodynamic interactions affects the polymer relaxation times, and in turn the polymer mean-square displacement (MSD). Analytically, an approximate expression of the MSD can be obtained  within the preaveraging approximation of the hydrodynamic tensor. In general, the  center-of-mass MSD is then given by \cite{mart:20,wink:20}
\begin{align} \label{eq:cm_msd}
\lla \Delta \bm r_{cm}^2 (t) \rra = \Lambda_p \frac{6 k_BT}{\gamma_T N_m} +   \lla \Delta \hat{\bm r}_{cm}^2 (t) \rra ,
\end{align}
with the contribution by activity:
\begin{align} \label{eq:cm_msd_abp}
\lla \Delta \hat{\bm r}_{cm}^2 (t) \rra  =  \Lambda_a \frac{2 v_0^2}{\gamma_R^2 N_m} \left( \gamma_R t -1 + e^{-\gamma_R t} \right) ,
\end{align}
where $\gamma_R = 2D_R$. The parameters $\Lambda_p$ and $\Lambda_a$ depend on the particular polymer environment and propulsion: D-ABPOs $\Lambda_p= \Lambda_a = 1$ \cite{eise:16}, for S-ABPOs $\Lambda_p = 1+ 3 \pi \eta \Omega_{00}$ and $\Lambda_a = 1$ \cite{mart:19}, and for E-ABPOs $\Lambda_p = \Lambda_a = 1+ 3 \pi \eta \Omega_{00}$ \cite{mart:20}, where  $\Omega_{00}$ is the preaveraged  Oseen tensor of the translational motion \cite{mart:19,mart:20,wink:20}.  Since $1+ 3 \pi \eta \Omega_{00} > 1$, hydrodynamic interactions accelerate the polymer dynamics.  The distinct $\Lambda_a$  values for S-ABPOs and E-ABPOs reflect the differences in the underlying propulsion mechanism --- in E-ABPOs,  $\Omega_{00}> 0$ is a consequence of the Stokeslets induced  by the active forces.

Our simulations of S-ABPOs yield polymer CM-MSDs, which, within the accuracy of the simulations,  are independent of polymer stiffness and hydrodynamic interactions for $Pe > 6$,  in agreement with the theoretical prediction, $\Lambda_a=1$. 

\begin{figure}[t]
 	\begin{center}
 		\includegraphics[width=1.0\columnwidth]{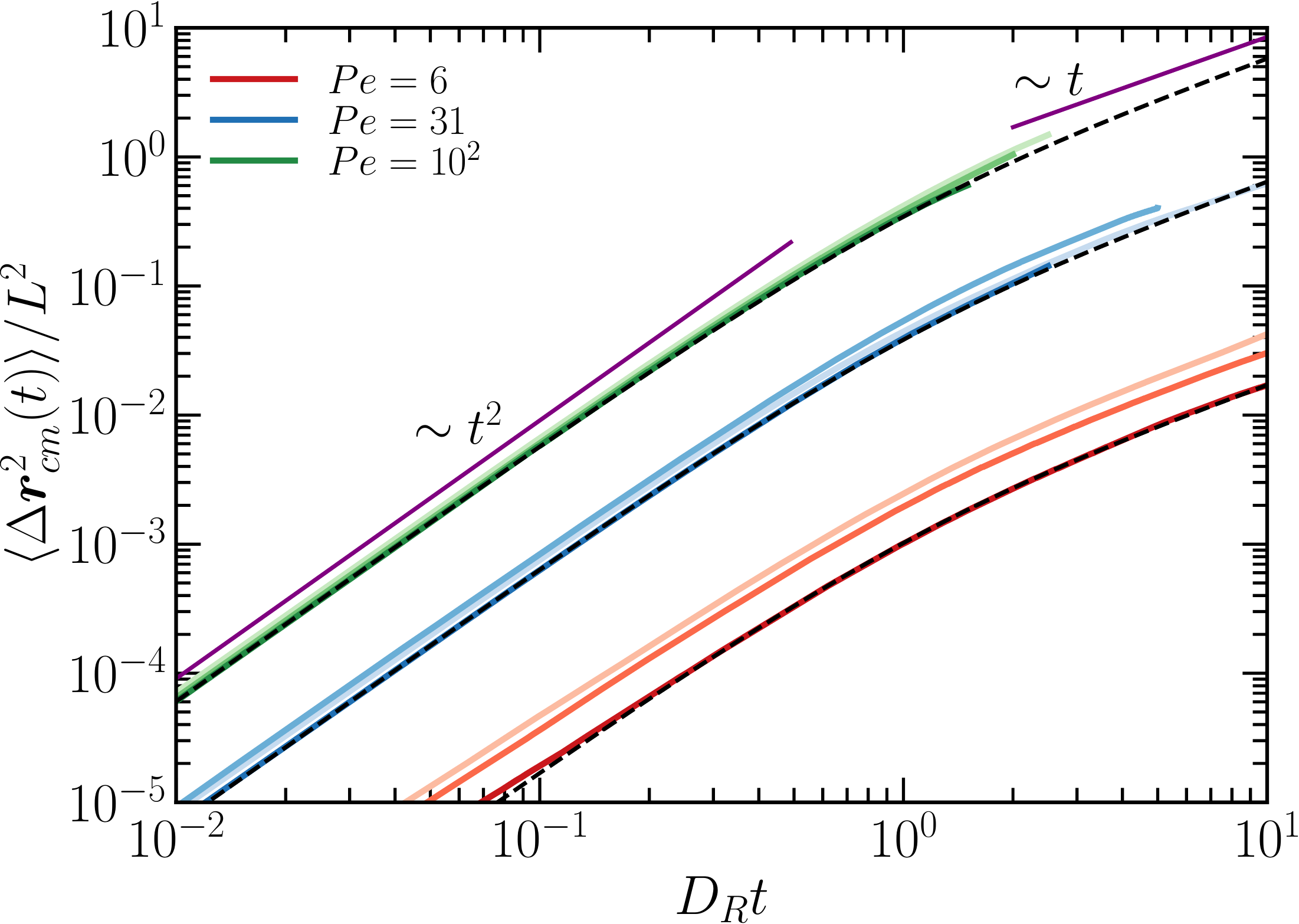}
 	\end{center}
 	\caption{Center-of-mass mean-square displacement of E-ABPOs  with $N_m=50$ monomers, the P\'eclet numbers $Pe=6$ (red, bottom), $31$ (blue, middle), and $100$ (green, top), and the bending rigidities $\kappa_b=0, 10,$ and  $10^3$ (bright to dark, top to bottom).  The dashed lines are examples of  fits of the CM-MSD in Eq.~\eqref{eq:cm_msd_abp} for $Pe=6$. The  fit parameters $\Lambda_a$ for all curves are  presented in Tab.~\ref{tab:hydr_param_ed}. The short purple lines present power laws with the indicated exponents.  }\label{fig:cm_msd_ed}
 \end{figure}

 Figure~\ref{fig:cm_msd_ed} depicts  MSDs for E-ABPOs together with fits to the  CM-MSD in Eq.~\eqref{eq:cm_msd_abp}, with $\Lambda_a$ as fit parameter. The simulation results follow the theoretical prediction very well in all time regimes --- the ballistic ($D_R t <1$) and the active diffusive regime ($D_R t > 1$). The fitting provides an estimation of the influence of hydrodynamic interactions on the MSD. As displayed in Tab.~\ref{tab:hydr_param_ed}, $\Lambda_a$ is in the range $7 < \Lambda_a < 22$, hence, the MSD is about an order of magnitude larger than that of S-ABPOs, where $\Lambda_a =1$. The $\Lambda_a$ values for $Pe=6$, where the polymers are still either rather flexible ($\kappa_b=0$) or stiff ($\kappa_b = 10, 10^3$), reflect the stronger effect of hydrodynamic interactions  for flexible polymers and its decreasing  influence with increasing bending rigidity (Tab.~\ref{tab:hydr_param_ed}) \cite{harn:96,petr:06}.  For $Pe=31$ and $Pe=100$, the polymer conformations are nearly independent of bending rigidity (Fig.~\ref{fig:end_SP_ED}), as reflected in the very similar $\Lambda_a$ values for these P\'eclet numbers, which vary by  approximately $20\%$, and are therefore within the accuracy of the simulations.

\setlength{\tabcolsep}{0.4em}
\begin{table} [b]
\caption{Hydrodynamic contribution $\Lambda_a$ to the active part of the center-of-mass  mean-square displacement (Eq.~\ref{eq:cm_msd_abp}) for various P\'eclet numbers, $Pe$, and bending  rigidities, $\kappa_b$, of E-ABPOs. }
\begin{tabular}{c |c c c | c c c | c c c  }
$Pe$ & \multicolumn{3}{|c|}{5}  & \multicolumn{3}{c|}{31} &  \multicolumn{3}{c}{100}     \\ \hline
$\kappa_b$  & 0 &  10 & 1000 &   0 &  10 & 1000 &  0 &  10 & 1000  \\
 $\Lambda_a$   & 21.5 &  15.9 & 9.0 &  10.0 &  12.2 & 8.8 & 9.2 &  8.3 & 7.6 \\
 \end{tabular}
\label{tab:hydr_param_ed} 
\end{table} 

\subsection{Dynamic structure factor}

Dynamical properties of polymers on various length scales  are accessible via the dynamic structure factor \cite{doi:86,harn:96}
\begin{align} \label{eq:dyn_struct_dis}
S(\bm q, t) =\frac{1}{N_m} \sum_{k=1}^{N_m} \sum_{n=1}^{N_m} \lla e^{-i \bm q \cdot (\bm r_k(t) - \bm r_n(0))} \rra .
\end{align} 
With the assumption of  a Gaussian distribution of the time-dependent monomer distance  $\Delta \bm r_{kn}(t) =\bm r_k(t) - \bm r_n(0)$, Eq.~\eqref{eq:dyn_struct_dis} becomes
\begin{align} \label{eq:dyn_struct_dis_gauss}
S(\bm q, t) =\frac{1}{N_m} \sum_{k=1}^{N_m} \sum_{n=1}^{N_m} \exp \left( - \frac{\bm q^2}{6} \lla \Delta \bm r^2_{kn}(t) \rra \right) .
\end{align} 
As is well known, for $qL \ll 1$, the polymer dynamics is determined by the center-of-mass motion, which yields  \cite{doi:86,harn:96}
\begin{align} \label{eq:dyn_struct_dis_cm}
S(\bm q, t) = S(\bm q) \exp \left( - \frac{\bm q^2}{6} \lla \Delta \bm r_{cm}^2(t) \rra \right) .
\end{align} 
 In contrast, on length scales $ ql \gg qL \gg 1$, the dynamic structure factor is determined by the polymer internal dynamics.

\subsubsection{Analytical consideration} \label{sec:dyn_struct_anal}

For a  continuous Gaussian (semiflexible) polymer model  \cite{wink:03,wink:94,bawe:85,lang:91,batt:87,ha:95}, the dynamic structure factor is given by \cite{harn:96}
\begin{align} \label{eq:dyn_struc_con_gauss}
S(\bm q, t) =\frac{1}{L^2} \int_0^L \int_0^L ds ds' \exp \left(  - \frac{\bm q^2}{6}  \lla \Delta \bm r^2 (s,s',t) \rra  \right) , 
\end{align} 
with the MSD $\langle  \Delta \bm r^2 (s,s',t) \rangle = \langle (\bm r(s,t) - \bm r(s',0))^2 \rangle$.  

An expression for the MSD $\langle  \Delta \bm r^2 (s,s',t) \rangle$ follows from the active Gaussian semiflexible polymer model presented in Refs.~\cite{eise:16,eise:17,mart:19,mart:20,wink:20}.  This model describes the properties of semiflexible ABPOs well for $Pe >10$, since the conformations depend only weakly on the persistence length for $pL >1$ (Fig.~\ref{fig:end_SP_ED}) \cite{eise:16,eise:17,mart:19,mart:20,wink:20}.  
The MSD  is then given by
\begin{align} \label{eq:displ_con} \nonumber 
 \lla  \Delta \bm r^2 (s,s',t) \rra =  & \lla \Delta \bm r_{cm}^2(t) \rra + \lla  \Delta \bm r^2 (s,s') \rra \\ &   + \lla  \Delta \bm r^2_{in} (s,s',t) \rra, 
\end{align}
with the intramolecular contribution 
\begin{align} \label{eq:dis_con_int} \nonumber 
 \lla  \Delta \bm r^2_{in} (s,s',t) \rra = &  \frac{4}{L} \sum_{m=1}^{\infty} \left( \lla \bm \chi_m^2(0) \rra - \lla 
 \bm \chi_m(t) \cdot \bm \chi_m(0)\rra\right) \\ &  \times \cos \left( \frac{m\pi s}{L} \right) \cos \left(\frac{m\pi s'}{L} \right) 
\end{align}
in terms of an eigenfunction representation, with the eigenfunctions $\cos(m \pi s/L$) for the mode $m \in \mathbb{N} \backslash 0$.  The $\bm \chi_m (t)$ are the normal-mode amplitudes.  Note that we neglect the coupling of different modes, which appears for systems with hydrodynamic interactions \cite{doi:86,harn:96}. Nevertheless, the expression captures the qualitative behavior \cite{hinc:09}. 
We focus on the effect of the internal polymer dynamics on the dynamic structure factor, with the relevant $q$-value range  $qL \gg 1$  and time scale $t/\tau_r \ll 1$, where $\tau_r$ denotes the longest polymer relaxation time. Then, the sum in Eq.~\eqref{eq:dis_con_int} is dominated by large  $m$ values, and the integrand has as a sharp peak at $s\approx s'$, hence, while evaluating the sum in Eq.~\eqref{eq:dis_con_int}, the   product of the eigenfunctions can be approximated  as $\cos (m\pi s/L) \cos (m\pi s'/L) \approx \cos (m\pi(s- s')/L) /2$ \cite{doi:86}.   As a consequence,
the displacement \eqref{eq:dis_con_int} is equal to half of the mean-square displacement of the point $\bm r(s,t)$ in the center-of-mass reference frame. The dynamic structure factor \eqref{eq:dyn_struc_con_gauss} is then given by 
\begin{align} \nonumber 
S(\bm q, t) = \frac{1}{L^2} & \int_0^L \int_0^L  ds ds' \exp \left( - \frac{\bm q^2}{6}  \lla \Delta \bm r^2 (s,s') \rra  \right) \\ & \times 
\exp \left( -  \frac{\bm q^2}{12} \lla  \Delta \bm r^2_{in} (s,t) \rra  \right)  .
\end{align}
The MSD of ABPOs has been discussed in detail in Refs.~\cite{eise:17,mart:19,mart:20} . For $Pe >10$,  $\gamma_R t = 2D_R t \ll 1$, and $t/\tau_r \ll 1$, the MSD is dominated by the activity-induced ballistic motion, which yields 
\begin{align}  \label{eq:int_dyn_act_t_quad}
 \lla  \Delta \bm r^2_{in} (s,t) \rra = &  \frac{v_0^2 l \gamma_R}{L} t^2   \sum_{m=1}^{\infty} \cos^2\left(\frac{m \pi s}{L}\right)  \Xi_m ,
 \end{align}
 with $\Xi_m =\tau_m/(1+\gamma_R \tau_m)$, $\tilde \tau_m/(1+\gamma_R  \tilde \tau_m)$, and $\tau_m^2/[\tilde \tau_m (1+\gamma_R \tilde \tau_m)]$ for D-ABPOs, S-ABPOs, and E-ABPOs, respectively, and   the relaxation times $\tau_m$ and  $\tilde \tau_m$ in absence or presence of hydrodynamic interactions, respectively.  Hence, in the active ballistic regime, the dynamic structure factor decays as 
 \begin{align} \label{eq:struc_fact_decay}
 S(\bm q,t) = S(\bm{q})  e^{- (\Gamma_q t)^{\zeta}} ,
 \end{align}
 with $\zeta = 2$ and a rate $\Gamma_q \sim q$.  In terms of the P\'eclet-number dependence,  $\Gamma_q \sim Pe^{4/3}$ for a D-ABPO. Since $\tau_m$  and $\tilde \tau_m$ depend on $Pe$, an analytical derivation of the $Pe$ dependence is difficult for S-ABPOs and E-ABPOs. 
 
 For long and flexible polymers, $pL \gg 1$, polymer characteristic dynamical regimes appear for $\gamma_R t >1$ and $t/\tau_r \ll 1$, where the MSD in the center-of-mass reference frame increases by a power law with an exponent smaller than unity \cite{eise:17,mart:19,mart:20}. Our  analytical studies predict the dependencies $ \lla  \Delta \bm r^2_{in} (s,t) \rra \sim Pe^{4/3} t^{1/2}$ (D-ABPO) \cite{eise:16}, $Pe^{3/2} t^{5/7}$ (E-APBO) \cite{mart:20}, and $Pe^{5/3} t^{1/3}$ (S-APBO) \cite{mart:19}. Hence, in this time regime,  the decay rate, $\Gamma_q$ in Eq.~\eqref{eq:struc_fact_decay},   exhibits the dependencies 
 $\Gamma_q \sim q^4 Pe^{8/3}$ (D-ABPO) --- this $q$ dependence is identical with that of a Rouse polymer ---, $\Gamma_q \sim  q^{14/5} Pe^{21/10}$ for E-ABPOs, and $\Gamma_q \sim  q^6 Pe^{5}$ for S-ABPOs. The latter indicate a decisive influence of hydrodynamic interactions om the distinct wave-number dependence.   
 
\begin{figure}[t]
  	\begin{center}
  		\includegraphics[width=1.0\columnwidth]{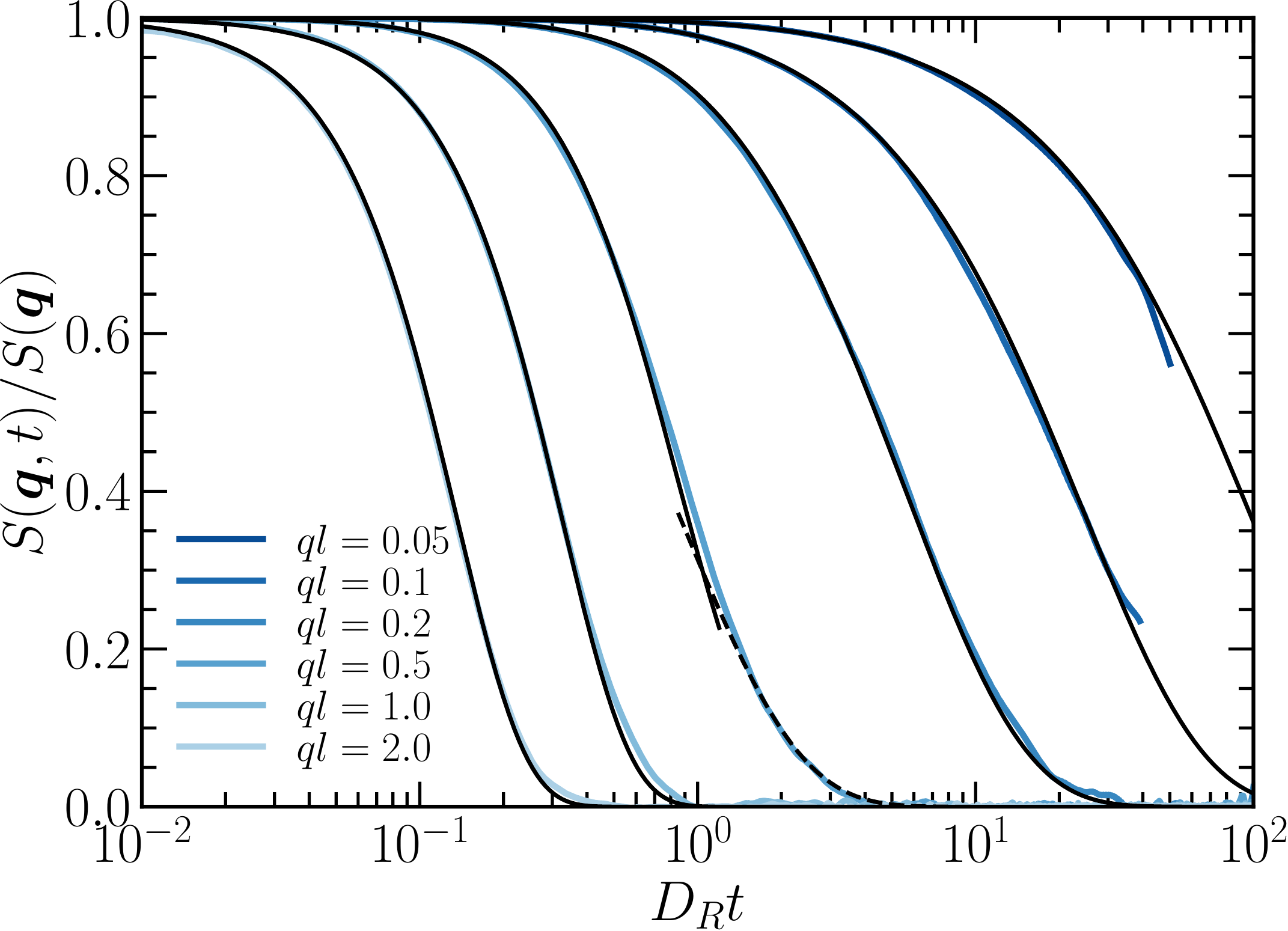}
  	\end{center}
  	\caption{Dynamic structure factors of S-ABPOs as a function of time for various $q$ values as indicated in the legend ($q$ values increase from right to left). The polymer stiffness is $\kappa_b=0$ and the P\'eclet number $Pe=31$. The blue lines indicate simulation results.  The (thin) black solid lines for $ql=0.05 - 0.5$ are obtained by Eq.~\eqref{eq:dyn_struct_dis_cm} with the CM-MSD of Eq.~\eqref{eq:cm_msd_abp}. The dashed line for $ql=0.5$ represents the linear dependence  $\ln(S(\bm q,t)) \sim - t$ and the (thin) black solid lines for $ql = 1.0, \, 2.0$ the power-law $\ln(S(\bm q,t))  \sim -  t^{1.75}$.} \label{fig:dyn_struct_sp_kb0_pe31}
\end{figure}

\begin{figure}[t]
   	\begin{center}
   		\includegraphics[width=1.0\columnwidth]{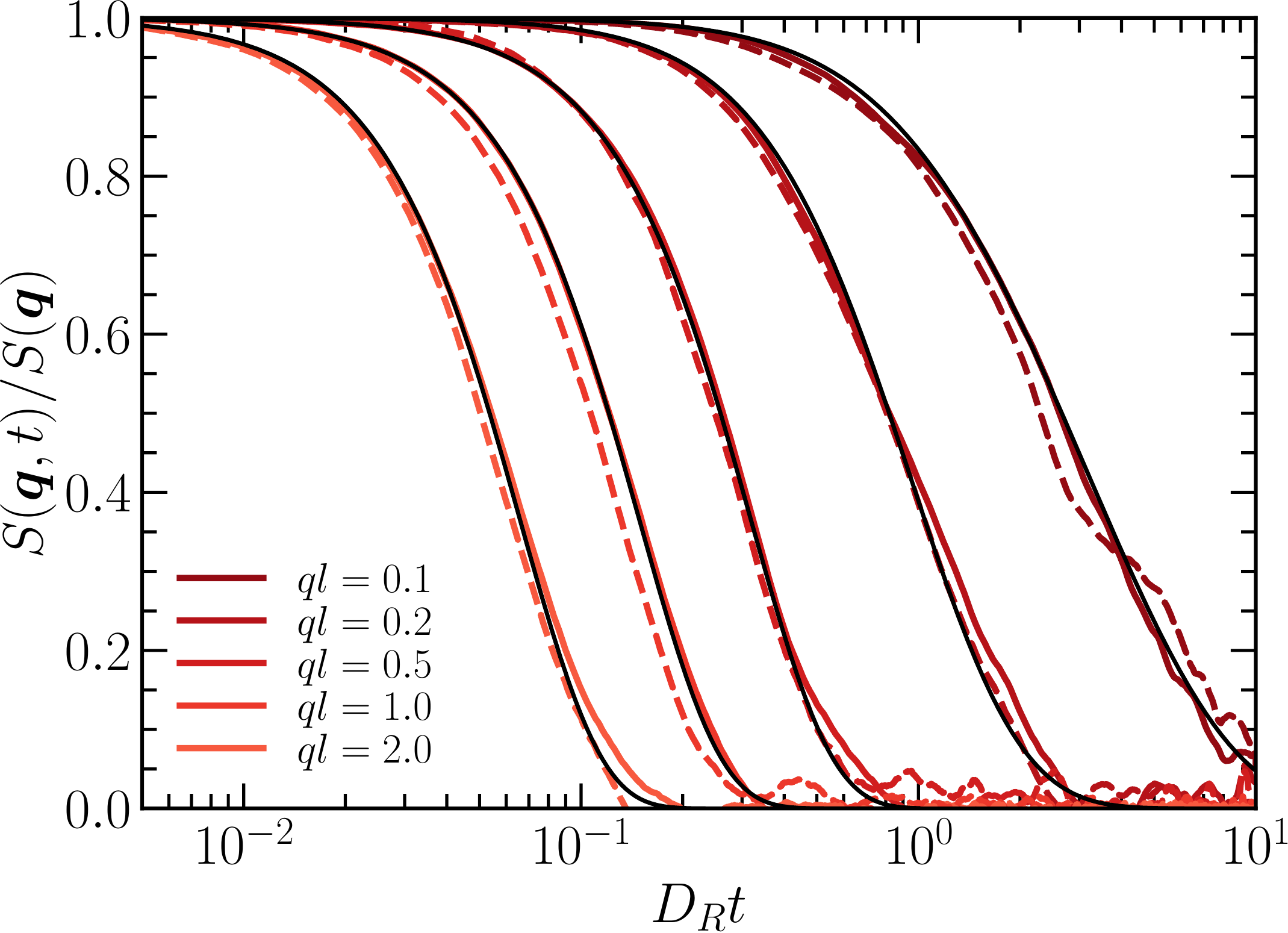}
   	\end{center}
   	\caption{Dynamic structure factors of E-ABPOs as a function of time for various $q$  values and the P\'eclet number $Pe=31$ ($q$ values increase from right to left). Simulation results for the polymer stiffness $\kappa_b=0$ are indicates by red solid lines and those for $\kappa_b=10$ by dashed red lines. The (thin) black solid lines for $ql=0.1, 0.2$ are obtained by Eq.~\eqref{eq:dyn_struct_dis_cm} with the CM-MSD of Eq.~\eqref{eq:cm_msd_abp};  for $ql \geq 0.5$, they represent the power-law $\ln(S(\bm q,t))  \sim -  t^{1.8}$.} \label{fig:dyn_struct_ed_pe31}
\end{figure}

\subsubsection{Simulation results}
 
Dynamic structure factors of S-ABPOs for the stiffness $\kappa_b=0$, the P\'eclet number $Pe=31$, and various $q$ values are displayed in  Figure~\ref{fig:dyn_struct_sp_kb0_pe31}. For $ql <0.5$,  the  $S(\bm q,t)$ curves are well described by Eq.~\eqref{eq:dyn_struct_dis_cm}, with $\langle \Delta \bm r_{cm}^2(t) \rangle$ of Eq.~\eqref{eq:cm_msd_abp}. This reveals the dominance of the  center-of-mass dynamics over the internal dynamics on this length- and time-scale.  However,  quantitative agreement between theory and simulation results is only achieved for factors $\Lambda_a$ larger than unity, namely  $\Lambda_a =1.28$ for $ql=0.05, \, 0.1$ and $\Lambda_a =1.4$ for $ql=0.2$. This could be due to limited  statistical accuracy of the simulation data, or, more likely, is a consequence of the applied approximations in the derivation of the analytical expression.  Similarly, the short-time behavior of the $S(\bm q,t)$ curve for $ql=0.5$ is well described by the center-of mass dynamics. However, for $D_R t >1$ $S(\bm q,t)$ is no longer determined by the CM-MSD, but still exhibits the dependence  $\ln (S(\bm q,t))  \sim - t$, characteristic for diffusion, yet with a different $\Gamma_q$ including contributions from the internal dynamics. This reflects the increasing importance of the internal polymer  dynamics on $S(\bm q,t)$ with increasing $q$ values.  For $q l  \geq 1.0$, the internal dynamics dominates the decay of the dynamic structure factor,  and $S(\bm q,t)$ is well described by Eq.~\eqref{eq:struc_fact_decay} with the exponent $\zeta \approx 1.75$.  Since $D_R t < 1$, the decay of $S(\bm q,t)$ occurs within the active  ballistic regime close to the crossover to the active diffusive regime (Fig.~\ref{fig:cm_msd_ed}), which explains that $\zeta$ is somewhat smaller than the theoretical value of a fully developed ballistic regime of $\zeta =2$.    Here, longer polymers are required to observe scaling with $q t$.  However, the power-law decay demonstrates the influence of activity on the internal dynamics, as has been discussed before in terms of the S-APBO mean-square displacement \cite{mart:19}. 

Figure~\ref{fig:dyn_struct_ed_pe31} presents dynamic structure factors of E-ABPOs for the bending parameters $\kappa_b=0$ and $\kappa_b=10$ for the activity $Pe=31$. As for the S-ABPOs,  for $ql  \leq 0.2$ $S(\bm q,t)$ is well described by the CM-MSD of the active polymer. Similarly, the structure factors for $ql \geq 0.5$ exhibit a power-law decay (Eq.~\eqref{eq:struc_fact_decay}), here with the exponent $\zeta \approx 1.8$, somewhat larger than that of S-ABPOs. The $S(\bm q,t)$ curves for the two persistence lengths are very similar, the  small horizontal shifts by up to $20\%$ are due to statistical inaccuracies. The polymers for the two stiffnesses exhibit the same dynamics on all length scales. This is not surprising, since the conformational properties are also nearly identical, as reflected by their mean-square end-to-end distances (Fig.~\ref{fig:end_SP_ED}) and the static structure factors (Fig.~\ref{fig:struct_kb}). Thus, the active forces dominate over the bending forces and determine the polymer conformational properties \cite{eise:16,eise:17,wink:20}. The exponent $\zeta  \approx 1.8$ is close to the value  $\zeta = 2$ for a ballistic active motion. The polymers are too short to exactly exhibit the active ballistic time dependence in the dynamic structure factor. 
 
 Considering the dynamic structure factors for the smaller activity of $Pe =6$, we find qualitatively the same behavior as displayed in Figs.~\ref{fig:dyn_struct_sp_kb0_pe31}, \ref{fig:dyn_struct_ed_pe31}. For $ql =0.1, 0.2$, $S(\bm q)$ is well described by Eq.~\eqref{eq:dyn_struct_dis_cm} with the CM-MSD of Eq.~\eqref{eq:cm_msd_abp}. Focusing on E-ABPOs, the respective curves for $\kappa_b=0$ and $\kappa_b=10$ are slight shifted with respect to each other, since the CM-MSD for $\kappa_b=0$ is slight larger at a given time $D_Rt$ (Fig.~\ref{fig:cm_msd_ed}). At $ql = 1, 2$, the curves for $\kappa_b=0$ and $\kappa_b=10$ are identical  within the numerical accuracy  with the exponent $\zeta \approx 1.6$, a value somewhat smaller than that for $Pe=31$.  As  in Fig.~\ref{fig:dyn_struct_sp_kb0_pe31}, for $ql=0.5$,  $S(\bm q)$ of E-ABPOs exhibits two time regimes,  where $\zeta \lesssim 2$ for shorter times and $\zeta \approx 1.3$ for longer times. Again, the latter reflects the increasing importance of the internal polymer dynamics with increasing $q$ values.

The dynamic structure factor of an anisotropic active Brownian particle (spherocylinder) has been determined theoretically and by simulations \cite{kurz:16}. Interestingly, $S(\bm q,t)$ exhibits damped oscillations over a certain range of wave numbers, which reflects the active persistent motion.  We observe oscillatory-type behavior with negative $S(\bm q,t)$ for $ql \gtrsim 1$ and $Pe =31$. However, we have not attempted to resolve it accurately. The oscillations are a particular feature of persistent (ballistic) active motion, but are not  necessary to characterize the ABP or ABPO dynamics.  The decay of  $S(\bm q,t)$  before the oscillations appear is already determined by the active persistent motion (Sec.~\ref{sec:dyn_struct_anal}).

\section{Summary and Conclusions} \label{sec:summary}

We have presented implementations of active polymers with either self-propelled monomers (S-ABPOs) or externally driven monomers (E-ABPOs) in a MPC fluid. In addition, we have analyzed their conformational and dynamical properties in terms of the static and dynamic structure factors.  

The force-free nature of S-ABPOs is captured in the  MPC  approach by a modification of the collision step, namely, the omission of  active velocities in the collision step and the consideration of the thermal velocities of the monomers only. The comparison of  mean square end-to-end distances obtained by MPC simulations  with results from Brownian dynamics simulations, accounting for hydrodynamic interactions via the Rotne-Prager-Yamakawa hydrodynamic tensor \cite{mart:19}, confirms the suitability of this approach.  

As previously reported \cite{mart:20,wink:20} and displayed in Fig.~\ref{fig:end_SP_ED}, the difference in the driving mechanism leads to substantially different polymer conformations, where S-ABPOs swell far less than E-ABPOs at high activities. This is reflected in the static structure factor, where even very stiff passive polymers exhibit a scaling behavior  with respect to the wave number, which deviates substantially from that of a rod on larger length scales. In contrast, even passive flexible E-ABPOs exhibit rod-like scaling for $Pe >30$ (Fig.~\ref{fig:struct_kb}). The observed dependencies are supported by analytical considerations. 

Simulation results for the active polymer center-of-mass mean-square displacement confirm the theoretical expectations \cite{mart:19,mart:20}, where the CM-MSD of S-ABPOs is independent of hydrodynamic contributions and agrees with that of D-ABPOs \cite{mart:19,wink:20}. However, the MSD of E-ABPOs is amplified by hydrodynamic interactions.  A fit of the theoretical expression Eq.~\eqref{eq:cm_msd_abp} provides a measure of the hydrodynamic enhancement, and shows that the center-of-mass dynamics of  E-ABPOs is  approximately an order of magnitude faster than that of S-ABPOs. The  polymer stiffness affects the contribution of  hydrodynamic interactions  for lower P\'eclet numbers ($Pe=6$), but the persistence-length dependence becomes weak for higher $Pe$, and thus for more extended polymers. This agrees with the behavior of passive semiflexible polymers, where hydrodynamic effects also become less important with increasing stiffness \cite{harn:96}. 

The dynamic structure factor reflects the overall active diffusive motion for small wave numbers $q$, and  the active internal dynamics for large $q$. Correspondingly, the time dependence of  $S(\bm q,t)$ is well described by the active center-of-mass mean-square displacement for $ql <0.5$. The nearly ballistic monomer motion leads to a stretched exponential decay of the dynamic structure factor for $ql \geq 0.5$, with an exponent $\zeta \approx 1.8$, smaller than the value $2$ of a ballistic motion. The  difference is a consequence of the shortness of the polymer of $N_m=50$ monomers.  Nevertheless, the dynamic structure factor reflects the active dynamics for S-ABPOs and E-ABPOs on all length scales.  

The outlined implementation of polymer hydrodynamics facilitates the study of more complex systems, where a Brownian dynamics simulation approach would fail, because of the lack of a suitable hydrodynamic tensor, e.g., for active objects in confinement or in complex geometries. Here, the MPC approach can provide the correct hydrodynamic behavior. The presented examples of active polymers serve as examples to confirm the suitability of our approach. The application to more challenging and interesting problems can be expected in the future. In particular, extensions are possible to include active stresses for a study of pusher/puller-type motile objects.  In Ref.~\cite{das:19.3}, a squirmer-type  colloidal particle has been implemented applying a similar strategy for the coupling of the colloid and the MPC fluid. Alternatively, motile point particles can be replaced by force dipoles, which again can be combined in a polymer.


%

\end{document}